\documentclass[
  aps,
  prd,
  reprint,
  notitlepage,
  amsmath,
  amssymb,
  superscriptaddress,
  nofootinbib
]{revtex4-2}

\usepackage{graphicx}
\usepackage{bm}
\usepackage{xcolor}
\newcommand{\revision}[1]{#1}

\begin{document}

\title{Validity Domains for Semiclassical Time in Quantum Cosmology}

\author{Gianpaolo Rosati}
\email{gianpaolo.rosati@polimi.it}
\affiliation{Dipartimento di Ingegneria Civile e Ambientale,
Politecnico di Milano, Piazza Leonardo da Vinci 32,
20133 Milano, Italy}

\author{Sergio L. Cacciatori}
\affiliation{Dipartimento di Scienza e Alta Tecnologia (DiSAT),
Universit\`a degli Studi dell'Insubria,
Via Valleggio 11, 22100 Como, Italy}
\affiliation{INFN, Sezione di Milano,
Via Celoria 16, 20133 Milano, Italy}
\affiliation{Como Lake Centre for AstroPhysics (CLAP),
DiSAT, Universit\`a degli Studi dell'Insubria,
Via Valleggio 11, 22100 Como, Italy}

\date{\today}

\begin{abstract}

We investigate the domain in which a semiclassical gravitational
WKB flow can serve as a quantitatively controlled temporal
reference for subordinate quantum degrees of freedom. Rather than
introducing WKB time at the outset, we formulate the
Born--Oppenheimer adiabatic condition directly on gravitational
configuration space using the Hamilton--Jacobi vector field. This
separates adiabatic control from the local WKB validity of the
gravitational branch and from the sharpness of the candidate
reference against quantum fluctuations. For a single selected
semiclassical branch, these requirements define the validity domain
\[
\mathcal D_t
=
\mathcal D_{\rm WKB}
\cap
\mathcal D_{\rm ad}
\cap
\mathcal D_{\rm sharp},
\]
with an additional decoherence condition when coherent branches
coexist.

In a closed Friedmann--Robertson--Walker minisuperspace coupled to a
homogeneous massive scalar mode, the distinction is explicit:
approaching the gravitational turning point from the oscillatory
region, the ordinary WKB control parameter diverges while the
configuration-space adiabatic parameter tends to zero. A uniform
Airy treatment of the turning point removes the formal WKB
divergence while preserving the inequivalence: the corresponding
local control parameter remains of order unity, whereas the
configuration-space adiabatic parameter is parametrically suppressed
as $O(M^{-1/3})$.

Thus Born--Oppenheimer adiabaticity and gravitational WKB validity
define inequivalent domains even within the simplest minisuperspace
model. We conclude that the existence of an oscillatory
semiclassical branch is not sufficient for a reliable WKB temporal
description: semiclassical temporal reliability is instead
determined by the overlap of the validity domains supporting the
reference flow.

\end{abstract}

\keywords{semiclassical time, quantum cosmology, Wheeler--DeWitt equation,
Born--Oppenheimer approximation, WKB approximation, gravitational turning points}

\maketitle

\section{Introduction}
In semiclassical quantum cosmology, an effective notion of time can
be recovered from the Wheeler--DeWitt equation by separating
gravitational and matter degrees of freedom and treating the
gravitational sector in a WKB approximation. The phase of a
semiclassical gravitational branch then defines a Hamilton--Jacobi
flow relative to which the subordinate quantum degrees of freedom
obey an effective Schr\"odinger dynamics
\cite{Banks1985,BroutVenturi1989,KieferSingh1991}. This construction and its
refinements, including backreaction, higher-order corrections, and
unitarity issues, have been extensively investigated and will not be
regarded here as new.
The existence of a semiclassical WKB branch, however, does not by
itself establish that the associated flow can serve as a controlled
temporal reference. Several logically distinct requirements enter
the semiclassical construction. The gravitational WKB approximation
must remain locally valid, transitions between the subordinate
Born--Oppenheimer channels must remain sufficiently suppressed, and
quantum fluctuations of the candidate reference sector must not
destroy the sharpness of the corresponding semiclassical flow.
When several WKB branches coexist, their effective autonomy raises
the additional issue of decoherence. These conditions are familiar
individually, but they need not define the same region of
configuration space.
This distinction is related to, but different from, several
developments in the semiclassical literature. Born--Oppenheimer
treatments of the matter--gravity system have addressed
backreaction, nonadiabatic corrections, and unitarity
\cite{BertoniFinelliVenturi1996,KieferSingh1991,
ManicciaMontaniAntonini2023}.
More recent
relational formulations have clarified the semiclassical expansion
and the status of the WKB time choice as a gauge-fixed relational
description \cite{Chataignier2022}. Recent work has also emphasized
the role of WKB structures in connecting quantum cosmological
equations with classical Friedmann dynamics
\cite{MatoneDimakis2025}. Decoherence, in turn, addresses
the effective autonomy of alternative semiclassical branches rather
than the local validity of the Hamilton--Jacobi flow itself
\cite{Halliwell1989}. Our question is distinct from these constructions: over which
region of gravitational configuration space are the conditions
required for a semiclassical flow to act as a controlled temporal
reference simultaneously satisfied?
This ordering is conceptually important because a validity condition
for the semiclassical separation should not presuppose the temporal
interpretation whose domain of applicability it is meant to assess.
We address this question by formulating the adiabatic control
condition directly on gravitational configuration space, before the
Hamilton--Jacobi flow is assigned a temporal interpretation. If
$S_0(Q)$ denotes the leading gravitational Hamilton--Jacobi
function, it defines the configuration-space vector field
\begin{equation}
V^A(Q)=G^{AB}\partial_B S_0 .
\end{equation}
The leading Born--Oppenheimer channel coupling can then be controlled
by a local dimensionless parameter constructed from $V^A$, the
instantaneous matter Hamiltonian, and the corresponding spectral
gaps, without first introducing $t_{\rm WKB}$. Only subsequently is
the derivative along the flow identified with
\begin{equation}
V^A\nabla_A
\equiv
\frac{\partial}{\partial t_{\rm WKB}} .
\end{equation}
Thus the adiabaticity condition required for the semiclassical
separation need not presuppose the temporal parametrization whose
domain of applicability is being tested.
A closed Friedmann--Robertson--Walker model with a homogeneous
massive scalar mode provides a simple discriminating test of this
construction. As the gravitational turning point is approached, the
ordinary WKB control parameter diverges, whereas the
configuration-space adiabatic parameter tends to zero. The
subordinate quantum dynamics therefore becomes locally more
adiabatic precisely where the gravitational WKB approximation
breaks down. This opposite behaviour demonstrates explicitly, within the model,
that the WKB and adiabatic validity domains are not interchangeable.
The minisuperspace model is used here as the simplest setting in
which this separation can be exhibited analytically, rather than as
a restriction of the validity-domain concept to homogeneous
cosmology.
These observations motivate a geometric characterization of the
domain in which a semiclassical gravitational flow can act as a
controlled temporal reference. For a single selected semiclassical
branch, we define
\begin{equation}
\mathcal D_t
=
\mathcal D_{\rm WKB}
\cap
\mathcal D_{\rm ad}
\cap
\mathcal D_{\rm sharp},
\end{equation}
where the three domains respectively control gravitational WKB
validity, Born--Oppenheimer adiabaticity, and the sharpness of
the candidate reference against quantum fluctuations. When several
semiclassical branches coexist coherently, effective branch
autonomy introduces an additional decoherence condition
$\mathcal D_{\rm dec}$. The construction therefore does not define
a new notion of time; it identifies where the usual semiclassical
WKB temporal parametrization is simultaneously supported by the
independent approximations required for its use as a physical
reference.
The paper is organized as follows. Section~II introduces the
semiclassical Wheeler--DeWitt framework and fixes the notation.
Section~III formulates Born--Oppenheimer adiabaticity directly on
gravitational configuration space. Section~IV defines the
independent validity domains entering the controlled temporal
reference. Section~V applies the construction to a closed FRW model
and analyzes the gravitational turning point. Section~VI discusses
the role of metrological thresholds, quantum fluctuations, and the
relation to existing semiclassical and relational approaches.
Section~VII summarizes the results and discusses their scope and
limitations.

\section{Semiclassical Wheeler--DeWitt framework}
We consider a constrained quantum system described by gravitational
configuration variables $Q^A$ and subordinate quantum degrees of
freedom $q$. The Wheeler--DeWitt equation is written schematically as

\begin{equation}
\left[
-\frac{\hbar^2}{2M}G^{AB}(Q)\nabla_A\nabla_B
+M U(Q)
+\hat H_q(Q)
\right]
\Psi(Q,q)
=
0,
\label{eq:WDW}
\end{equation}

where $G^{AB}$ is the inverse metric on gravitational configuration
space, $U(Q)$ is the gravitational potential, $\hat H_q(Q)$ is the
Hamiltonian of the subordinate sector, and $M$ denotes the large
gravitational scale controlling the semiclassical expansion.
At each point $Q$ of gravitational configuration space, let
$\{|n;Q\rangle\}$ denote an instantaneous orthonormal basis of the
subordinate sector satisfying

\begin{equation}
\hat H_q(Q)|n;Q\rangle
=
\varepsilon_n(Q)|n;Q\rangle .
\label{eq:instantaneous_basis}
\end{equation}

The total wave function may then be expanded as

\begin{equation}
\Psi(Q,q)
=
\sum_n
\chi_n(Q)\,
\langle q|n;Q\rangle .
\label{eq:BO_expansion}
\end{equation}

The dependence of the instantaneous states on $Q$ generates
off-diagonal derivative couplings between different channels. These
couplings will provide the configuration-space formulation of
adiabatic control used below.
In the semiclassical regime, each gravitational channel is written
in WKB form as

\begin{equation}
\chi_n(Q)
=
A_n(Q)
\exp\left[
\frac{iM}{\hbar}S_0(Q)
\right]
\psi_n(Q),
\label{eq:WKB_ansatz}
\end{equation}

where $S_0(Q)$ is the leading gravitational phase and $A_n(Q)$
contains the corresponding semiclassical amplitude. At leading
order in the large gravitational scale $M$, insertion into
Eq.~\eqref{eq:WDW} yields the Hamilton--Jacobi equation

\begin{equation}
\frac{1}{2}
G^{AB}
\partial_A S_0
\partial_B S_0
+
U(Q)
=
0.
\label{eq:HJ}
\end{equation}

The Hamilton--Jacobi function therefore defines a vector field on
gravitational configuration space,

\begin{equation}
V^A(Q)
\equiv
G^{AB}(Q)\partial_B S_0(Q).
\label{eq:HJ_flow}
\end{equation}

At this stage $V^A$ is introduced purely as a configuration-space
flow. No temporal interpretation is required.

\section{Configuration-space adiabaticity before time}

The dependence of the instantaneous basis on the gravitational
variables produces derivative couplings between different
Born--Oppenheimer channels. At leading semiclassical order, the
factor $M$ carried by the derivative of the WKB phase cancels the
$1/M$ multiplying the gravitational kinetic operator in
Eq.~\eqref{eq:WDW}. The resulting leading off-diagonal coupling
between two instantaneous states $|n;Q\rangle$ and $|m;Q\rangle$
is therefore

\begin{equation}
C_{mn}^{(0)}
=
-i\hbar\,
V^A
\langle m;Q|\nabla_A n;Q\rangle .
\label{eq:leading_channel_coupling}
\end{equation}

For nondegenerate instantaneous eigenvalues, differentiation of
Eq.~\eqref{eq:instantaneous_basis} gives

\begin{equation}
\langle m;Q|\nabla_A n;Q\rangle
=
\frac{
\langle m;Q|
\nabla_A\hat H_q
|n;Q\rangle
}{
\varepsilon_n-\varepsilon_m
},
\qquad
m\neq n .
\label{eq:eigenstate_derivative}
\end{equation}

Hence

\begin{equation}
C_{mn}^{(0)}
=
-i\hbar\,
\frac{
V^A
\langle m;Q|
\nabla_A\hat H_q
|n;Q\rangle
}{
\varepsilon_n-\varepsilon_m
},
\qquad
m\neq n .
\label{eq:channel_coupling_H}
\end{equation}

All quantities entering this expression are defined on
gravitational configuration space. In particular, no temporal
parameter has yet been introduced.
The normalization of $V^A$ is understood to be fixed by the
Hamiltonian constraint and lapse convention adopted in the
semiclassical reduction. The configuration-space formulation does
not remove this parametrization freedom; rather, it allows
adiabatic control to be assessed before the corresponding flow
parameter is interpreted as $t_{\rm WKB}$.
Adiabatic separation requires the off-diagonal channel coupling to
remain small compared with the corresponding instantaneous spectral
gap. This motivates the local dimensionless parameter

\begin{equation}
\boxed{
\epsilon_{mn}^{\rm cfg}(Q)
\equiv
\hbar
\frac{
\left|
V^A(Q)
\langle m;Q|
\nabla_A\hat H_q(Q)
|n;Q\rangle
\right|
}{
\left|
\varepsilon_n(Q)-\varepsilon_m(Q)
\right|^2
},
\qquad
m\neq n .
}
\label{eq:cfg_adiabatic_parameter}
\end{equation}

The Born--Oppenheimer channel separation is locally controlled when

\begin{equation}
\epsilon_{mn}^{\rm cfg}(Q)\ll1
\label{eq:cfg_adiabatic_condition}
\end{equation}

for the transitions relevant to the state under consideration.
Crucially, Eq.~\eqref{eq:cfg_adiabatic_parameter} is formulated
entirely in terms of the Hamilton--Jacobi flow, the instantaneous
subordinate Hamiltonian, and its spectral gaps. Its definition
therefore does not require a prior identification of $V^A$ with a
time derivative.
Only after the configuration-space flow and its adiabatic control
have been defined do we introduce the usual semiclassical temporal
parametrization. Along a Hamilton--Jacobi trajectory, we define

\begin{equation}
\frac{\partial}{\partial t_{\rm WKB}}
\equiv
V^A\nabla_A .
\label{eq:WKB_time}
\end{equation}

With this identification, Eq.~\eqref{eq:cfg_adiabatic_parameter}
takes the familiar time-dependent form

\begin{equation}
\epsilon_{mn}^{\rm cfg}
=
\hbar
\frac{
\left|
\langle m|
\partial_{t_{\rm WKB}}\hat H_q
|n\rangle
\right|
}{
\left|
\varepsilon_n-\varepsilon_m
\right|^2
},
\qquad
m\neq n .
\label{eq:temporal_adiabatic_parameter}
\end{equation}

The temporal adiabatic criterion is therefore recovered as the
representation along a Hamilton--Jacobi trajectory of a condition
that was already defined on gravitational configuration space.

\section{Independent validity domains for a temporal reference}
The existence of a Hamilton--Jacobi flow does not by itself imply
that the WKB approximation supporting it is locally controlled.
For a semiclassical gravitational branch of the form

\begin{equation}
\chi(Q)
=
A(Q)
\exp\left[
\frac{iM}{\hbar}S_0(Q)
\right],
\end{equation}

the WKB approximation requires the amplitude to vary slowly on the
local phase scale. We denote by

\begin{equation}
\delta_{\rm WKB}(Q)
\end{equation}

a dimensionless local measure of the corresponding higher-order
WKB corrections, and define the controlled WKB domain as

\begin{equation}
\boxed{
\mathcal D_{\rm WKB}
=
\left\{
Q:
\delta_{\rm WKB}(Q)
<
\kappa_{\rm WKB}
\right\},
}
\label{eq:DWKB}
\end{equation}

where $\kappa_{\rm WKB}\ll1$ specifies the required level of
semiclassical accuracy.
The adiabatic condition is controlled independently by the
configuration-space parameter introduced in
Eq.~\eqref{eq:cfg_adiabatic_parameter}. For a specified set
$\mathcal T$ of transitions relevant to the subordinate state, we
define

\begin{equation}
\epsilon_{\rm ad}(Q)
\equiv
\max_{(m,n)\in\mathcal T}
\epsilon_{mn}^{\rm cfg}(Q),
\label{eq:epsilon_ad}
\end{equation}

and the corresponding adiabatic domain as

\begin{equation}
\boxed{
\mathcal D_{\rm ad}
=
\left\{
Q:
\epsilon_{\rm ad}(Q)
<
\kappa_{\rm ad}
\right\},
}
\label{eq:Dad}
\end{equation}

where $\kappa_{\rm ad}\ll1$ specifies the required suppression of
nonadiabatic channel mixing.

The definition of $\mathcal D_{\rm ad}$ is logically distinct from
the WKB condition: it tests the separation of the subordinate
instantaneous channels along the Hamilton--Jacobi flow, rather than
the local validity of the gravitational WKB approximation itself.
A third condition concerns the sharpness of the candidate
semiclassical reference. Even when the mean Hamilton--Jacobi flow is
well defined, quantum fluctuations of the subordinate sector may
induce fluctuations in the gravitational flow and thereby limit its
interpretation as a single sharply defined reference.

We denote by

\begin{equation}
\eta_{\rm fluc}(Q)
\end{equation}

a dimensionless measure of the fluctuations relevant to the
candidate reference, and define

\begin{equation}
\boxed{
\mathcal D_{\rm sharp}
=
\left\{
Q:
\eta_{\rm fluc}(Q)
<
\kappa_{\rm fluc}
\right\},
}
\label{eq:Dsharp}
\end{equation}

where $\kappa_{\rm fluc}\ll1$ specifies the required sharpness of the
semiclassical reference.

This condition is distinct from mean backreaction. A large mean
correction may require a self-consistent redefinition of the
semiclassical background, whereas large fluctuations directly limit
the use of a single sharply defined flow as a common temporal
reference.
For a single semiclassical branch that has already been selected,
the controlled temporal domain is defined by the simultaneous
validity of the three independent conditions,

\begin{equation}
\boxed{
\mathcal D_t
=
\mathcal D_{\rm WKB}
\cap
\mathcal D_{\rm ad}
\cap
\mathcal D_{\rm sharp}.
}
\label{eq:Dt}
\end{equation}

Membership in only one or two of these domains is not sufficient.
The existence of a Hamilton--Jacobi flow, adiabatic suppression of
channel transitions, and reference sharpness test different aspects
of the semiclassical construction and must therefore be controlled
separately.
When several semiclassical branches coexist coherently, an
additional requirement concerns their effective autonomy. Denoting
by $\mathcal D_{\rm dec}$ the region in which interference between
the relevant branches is sufficiently suppressed, the corresponding
multibranch domain is

\begin{equation}
\boxed{
\mathcal D_t^{\rm multi}
=
\mathcal D_{\rm WKB}
\cap
\mathcal D_{\rm ad}
\cap
\mathcal D_{\rm sharp}
\cap
\mathcal D_{\rm dec}.
}
\label{eq:Dt_multi}
\end{equation}

The decoherence condition is therefore conditional rather than
universal: it is required when several candidate reference branches
coexist, but it does not constitute an additional constraint when a
single branch has already been selected.
The thresholds
$\kappa_{\rm WKB}$, $\kappa_{\rm ad}$, and $\kappa_{\rm fluc}$
should not be interpreted at this stage as new fundamental
constants. They specify how accurately the corresponding
approximation or reference property is required to be controlled.
The structural content of the construction lies instead in the
existence of distinct validity conditions and in the geometry of
their overlap in configuration space.

The dependence of the domain boundaries on these thresholds, and
their interpretation in terms of metrological accuracy, will be
discussed in Sec.~VI.

\section{Closed FRW universe as a discriminating test}
We now test the independence of the WKB and adiabatic validity
conditions in the simplest explicit setting considered here: a
closed Friedmann--Robertson--Walker minisuperspace with scale factor
$a$ and a homogeneous massive scalar mode $\phi$.
We adopt a normalization in which the large gravitational scale
$M$ has been factored out of the leading Hamilton--Jacobi equation
and remains explicit in the WKB phase
$\exp(iMS_0/\hbar)$.
The gravitational Hamilton--Jacobi equation can be written in the
form

\begin{equation}
-\frac{1}{12a}
\left(
\frac{dS_0}{da}
\right)^2
-3a
+3H_\Lambda^2a^3
=
0,
\label{eq:FRW_HJ}
\end{equation}

or equivalently,

\begin{equation}
\left(
\frac{dS_0}{da}
\right)^2
=
36a^2
\left(
H_\Lambda^2a^2-1
\right).
\label{eq:FRW_HJ_squared}
\end{equation}
The oscillatory WKB branch therefore exists in the
classically allowed region

\begin{equation}
a>a_*,
\qquad
a_*
=
H_\Lambda^{-1},
\label{eq:FRW_turning_point}
\end{equation}

where $a_*$ is the gravitational turning point.

The purpose of this model is not to identify the turning point with
the boundary of semiclassical time. It is instead to compare,
within the same configuration-space region, the behaviour of the
independent WKB and Born--Oppenheimer control parameters.
For the oscillatory gravitational branch, define the local
Hamilton--Jacobi momentum

\begin{equation}
p_a(a)
\equiv
\frac{dS_0}{da}
=
-6a\sqrt{H_\Lambda^2a^2-1},
\label{eq:FRW_pa}
\end{equation}
Since the inverse minisuperspace metric associated with
Eq.~\eqref{eq:FRW_HJ} is

\begin{equation}
G^{aa}
=
-\frac{1}{6a},
\label{eq:FRW_supermetric}
\end{equation}

the expanding branch is selected by the negative sign of $p_a$.
The corresponding Hamilton--Jacobi flow is therefore
\begin{equation}
V^a(a)
=
\sqrt{H_\Lambda^2a^2-1}.
\label{eq:FRW_flow}
\end{equation}

A standard local measure of WKB control is

\begin{equation}
\delta_{\rm WKB}(a)
\equiv
\frac{\hbar}{M}
\left|
\frac{p_a'(a)}{p_a^2(a)}
\right|,
\label{eq:FRW_deltaWKB}
\end{equation}

which provides the local WKB control measure used in the following
turning-point analysis.

Using Eq.~\eqref{eq:FRW_pa}, one finds

\begin{equation}
p_a(a)\longrightarrow0
\qquad
\text{for}
\qquad
a\longrightarrow a_*^+,
\end{equation}

while $p_a'(a)$ becomes singular. Consequently,

\begin{equation}
\boxed{
\delta_{\rm WKB}(a)
\longrightarrow
\infty
\qquad
\text{for}
\qquad
a\longrightarrow a_*^+ .
}
\label{eq:FRW_WKB_breakdown}
\end{equation}

The ordinary gravitational WKB approximation therefore loses local
control as the turning point is approached from the oscillatory
region.
The subordinate sector is taken to be a homogeneous massive scalar
mode. Its instantaneous Hamiltonian is

\begin{equation}
\hat H_\phi(a)
=
\frac{\hat p_\phi^2}{2a^3}
+
\frac{1}{2}a^3m^2\hat\phi^2 .
\label{eq:scalar_H}
\end{equation}

At fixed $a$, this is a harmonic oscillator with effective mass
$a^3$ and frequency $m$. Its instantaneous eigenvalues are therefore

\begin{equation}
\varepsilon_n
=
\hbar m
\left(
n+\frac{1}{2}
\right),
\label{eq:scalar_spectrum}
\end{equation}

which are independent of $a$, although the corresponding
eigenstates retain an explicit $a$ dependence.

The variation of the instantaneous basis with $a$ generates
nonadiabatic transitions. Since the $a$ dependence enters through
the oscillator width, the leading off-diagonal couplings connect
states differing by two oscillator quanta,

\begin{equation}
n
\longleftrightarrow
n\pm2 .
\label{eq:scalar_selection_rule}
\end{equation}
Using the instantaneous oscillator ladder operators, one finds

\begin{equation}
\left|
\left\langle n+2;a
\left|
\partial_a\hat H_\phi
\right|
n;a
\right\rangle
\right|
=
\frac{3\hbar m}{2a}
\sqrt{(n+1)(n+2)} .
\label{eq:scalar_matrix_element}
\end{equation}

Since

\begin{equation}
\left|
\varepsilon_{n+2}-\varepsilon_n
\right|
=
2\hbar m,
\label{eq:scalar_gap}
\end{equation}

Substituting Eqs.~\eqref{eq:FRW_flow},
\eqref{eq:scalar_matrix_element}, and
\eqref{eq:scalar_gap} into
Eq.~\eqref{eq:cfg_adiabatic_parameter} gives

\begin{equation}
\epsilon_{n+2,n}^{\rm cfg}(a)
=
\frac{3}{8}
\frac{H_\Lambda}{m}
\sqrt{
1-\frac{1}{H_\Lambda^2a^2}
}
\sqrt{(n+1)(n+2)} .
\label{eq:FRW_epsilon_ad}
\end{equation}

Approaching the gravitational turning point from the oscillatory
region,

\begin{equation}
a\longrightarrow a_*^+,
\end{equation}

the Hamilton--Jacobi flow vanishes and therefore

\begin{equation}
\boxed{
\epsilon_{n+2,n}^{\rm cfg}(a)
\longrightarrow0 .
}
\label{eq:FRW_adiabatic_limit}
\end{equation}
Combining Eqs.~\eqref{eq:FRW_WKB_breakdown} and
\eqref{eq:FRW_adiabatic_limit}, one obtains

\begin{equation}
\boxed{
\delta_{\rm WKB}(a)\longrightarrow\infty,
\qquad
\epsilon_{n+2,n}^{\rm cfg}(a)\longrightarrow0,
\qquad
a\longrightarrow a_*^+ .
}
\label{eq:FRW_opposite_limits}
\end{equation}
For any fixed positive control thresholds
$\kappa_{\rm WKB}$ and $\kappa_{\rm ad}$, there is therefore a
sufficiently small neighbourhood of the turning point in which the
adiabatic condition is satisfied while the gravitational WKB
condition is violated.
Thus the subordinate quantum dynamics becomes locally more
adiabatic precisely where the gravitational WKB approximation loses
control. Neither limiting behaviour is by itself unexpected: WKB breakdown
near a turning point is standard, while the local adiabatic coupling
is suppressed as the Hamilton--Jacobi flow tends to zero. The
relevant result is their simultaneous occurrence within the same
configuration-space region. The vanishing of the adiabatic
parameter is therefore not interpreted as a restoration of the
semiclassical clock near the turning point. Rather, it demonstrates
that increasingly accurate Born--Oppenheimer channel separation can
coexist with increasingly poor WKB control of the gravitational
reference that generates the flow.

Consequently,

\begin{equation}
\boxed{
\mathcal D_{\rm ad}
\neq
\mathcal D_{\rm WKB}.
}
\end{equation}
Born--Oppenheimer adiabaticity is therefore not sufficient to
establish the validity of the semiclassical temporal reference.
The gravitational turning point therefore has two distinct roles
that should not be conflated. It marks the boundary between the
oscillatory and nonoscillatory character of the leading
Wheeler--DeWitt solution, but it does not by itself determine the
boundary of a controlled semiclassical temporal description.

In particular, the existence of a real Hamilton--Jacobi flow for

\begin{equation}
a>a_*
\end{equation}

is only a necessary starting point. A controlled temporal reference
requires, in addition, that the corresponding configuration lie
inside the overlap of the relevant validity domains. Thus,

\begin{equation}
\boxed{
\begin{gathered}
\text{oscillatory semiclassical branch}\\
\not\Rightarrow\\
\text{controlled temporal reference}
\end{gathered}
}
\label{eq:oscillatory_not_controlled}
\end{equation}

The closed FRW model therefore provides an explicit counterexample
to the identification of Born--Oppenheimer adiabaticity with
gravitational WKB validity. Its role in the present analysis is
discriminating rather than universal: it establishes that the two
conditions can define different domains even within the simplest
minisuperspace setting.

The result does not imply that the turning point is a fundamental
boundary for every semiclassical clock, nor does it establish that
the FRW scale factor is a universally preferred temporal variable.
It shows instead that the validity of a semiclassical WKB temporal
reference cannot, in general, be inferred from adiabatic control of
the subordinate sector alone.
The opposite turning-point behavior found above is not an artifact
of the singularity of the ordinary WKB approximation.  A uniform
Airy treatment of the gravitational turning point, presented in
Appendix~\ref{app:airy}, removes the formal divergence of the local
WKB indicator but does not restore equivalence between the two
control conditions.  At the turning point one finds
\begin{equation}
\delta_A(a_*)=\frac{2}{\sqrt{3}}=O(1),
\qquad
\epsilon_{\rm cfg}^{A}(a_*)=O(M^{-1/3}),
\end{equation}
so that in the semiclassical regime the subordinate dynamics remains
parametrically adiabatic while the gravitational WKB description is
not parametrically controlled.

\section{Metrological thresholds, fluctuations, and relation to existing approaches}
The validity domains introduced in Sec.~IV depend on the thresholds
$\kappa_{\rm WKB}$, $\kappa_{\rm ad}$, and $\kappa_{\rm fluc}$.
These parameters do not introduce new dynamical scales. They specify
the accuracy required before a given approximation, or a given
reference property, is regarded as sufficiently controlled for the
intended use.

Accordingly, a change in a threshold changes the quantitative
location of the corresponding domain boundary,

\begin{equation}
\delta_{\rm WKB}=\kappa_{\rm WKB},
\qquad
\epsilon_{\rm ad}=\kappa_{\rm ad},
\qquad
\eta_{\rm fluc}=\kappa_{\rm fluc},
\end{equation}

but does not alter the logical distinction between the criteria
themselves. In this sense, the thresholds have a metrological role:
they translate a required level of accuracy into a boundary in
configuration space.
Threshold dependence should therefore be distinguished from
arbitrariness in the physical criterion. The quantities
$\delta_{\rm WKB}$, $\epsilon_{\rm ad}$, and $\eta_{\rm fluc}$
probe different properties of the semiclassical construction
independently of the numerical tolerances chosen to delimit their
controlled regimes. The thresholds determine how strictly those
properties are required to be satisfied, not which properties must
be tested.

This distinction is already visible in the FRW example. Changing $\kappa_{\rm WKB}$ or $\kappa_{\rm ad}$ shifts the
quantitative boundaries of the corresponding domains, but, within
the ordinary WKB treatment, it cannot remove the opposite
turning-point behaviour
\begin{equation}
\delta_{\rm WKB}\longrightarrow\infty,
\qquad
\epsilon_{\rm cfg}^{(n+2,n)}\longrightarrow0.
\end{equation}

The non-equivalence of the two control conditions is therefore
structural within the model, whereas the precise location of a
chosen validity boundary is accuracy dependent.
The sharpness condition requires a distinction between mean
backreaction and quantum fluctuations. At subleading semiclassical
order, the expectation value of the subordinate Hamiltonian modifies
the mean gravitational Hamilton--Jacobi flow. If this correction is
large, the original unbackreacted branch may cease to provide an
accurate approximation, but a self-consistent semiclassical branch
including the mean backreaction may still exist.

Quantum fluctuations have a different status. Writing

\begin{equation}
\Delta H_q
=
\sqrt{
\langle \hat H_q^2\rangle
-
\langle \hat H_q\rangle^2
},
\label{eq:DeltaHq}
\end{equation}

the dispersion of the subordinate energy provides, in the present
minisuperspace setting, a proxy for the dispersion induced in the
subleading correction to the gravitational reference flow. The
quantity $\eta_{\rm fluc}$ introduced in Sec.~IV should therefore be
understood as a measure of reference sharpness rather than merely of
the magnitude of the mean matter correction.
The use of $\Delta H_q$ in this construction is deliberately
limited to the minisuperspace model. In semiclassical gravity, the
role of stress--energy fluctuations as a consistency condition has
long been recognized \cite{KuoFord1993}. A purely local comparison
between the variance and the mean stress tensor, however, is not in
general sufficient to diagnose the validity of the semiclassical
description; the result depends on the physical resolution or
smearing scale at which the fluctuations are probed
\cite{PhillipsHu2000}.

A field-theoretic extension of $\mathcal D_{\rm sharp}$ would
therefore require appropriately renormalized stress--energy
correlation functions, or an equivalent measure of the metric
fluctuations induced around the mean semiclassical geometry.
In a full quantum field theory, local stress--energy fluctuations
are ultraviolet singular and should instead be characterized through
a smeared operator,
\begin{equation}
T_{\mu\nu}(f)
=
\int d^4x\,\sqrt{-g}\,
f(x)\,T_{\mu\nu}(x),
\end{equation}
where the physical support and characteristic scale of the test
function $f(x)$ specify the spacetime region over which the
fluctuations are averaged. In the present validity-domain language,
this smearing scale would acquire a metrological interpretation:
it fixes the physical resolution at which the sharpness of the
candidate semiclassical reference is assessed.

We do not attempt such a field-theoretic extension here.
Accordingly, $\Delta H_q$ and the associated $\eta_{\rm fluc}$
should be regarded as minisuperspace proxies for the more general
requirement that the candidate gravitational reference remain
sufficiently sharp.
Decoherence addresses a different consistency problem. In a state
containing several semiclassical WKB components, suppression of
interference is required before the individual branches can be
treated as effectively autonomous alternatives
\cite{Halliwell1989}. This requirement concerns branch autonomy
rather than the local WKB or Born--Oppenheimer validity of a given
branch.

Accordingly, the existence of semiclassical branches and their
effective autonomy should be distinguished,

\begin{equation}
\begin{gathered}
\text{existence of WKB branches}\\
\neq\\
\text{effective autonomy of WKB branches}
\end{gathered}
\end{equation}

This is why $\mathcal D_{\rm dec}$ enters the present construction
only conditionally. For a single branch already selected, for
example by boundary conditions in quantum cosmology
\cite{Vilenkin1986} or by the physical preparation of the state,
the controlled temporal domain remains $\mathcal D_t$ as defined in
Eq.~\eqref{eq:Dt}.
Relational approaches address the absence of an external time
parameter from a complementary perspective. In the Page--Wootters
construction, evolution is recovered through conditional relations
between a clock subsystem and the remaining degrees of freedom
\cite{PageWootters1983}. More recent formulations have connected
semiclassical WKB time with relational observables and gauge fixing
in constrained quantum systems \cite{Chataignier2022}.

The present construction addresses a different question. It does
not propose an alternative relational clock, nor does it assign
fundamental status to $t_{\rm WKB}$. Instead, it asks under which
conditions a semiclassical gravitational flow that has already been
identified as a candidate reference remains quantitatively
controlled. In this sense, $\mathcal D_t$ is a domain of
applicability of the reference description rather than a new
definition of relational time.
The construction should also be distinguished from the traditional
semiclassical program itself. The recovery of an effective
Schr\"odinger dynamics from the Wheeler--DeWitt equation, together
with systematic treatments of Born--Oppenheimer separation,
backreaction, higher-order corrections, and unitarity, has a long
history \cite{Banks1985,BroutVenturi1989,KieferSingh1991,
BertoniFinelliVenturi1996,Kiefer2012}.
None of these ingredients is claimed here as new.

The present contribution is instead to organize their consistency
requirements around the physical role assigned to the
semiclassical gravitational sector as a temporal reference. In
particular, the Born--Oppenheimer adiabatic condition is formulated
on gravitational configuration space before the Hamilton--Jacobi
flow is interpreted temporally, and its validity domain is then
kept distinct from gravitational WKB control and reference
sharpness. The FRW example demonstrates explicitly that WKB validity and
Born--Oppenheimer adiabaticity cannot in general be replaced by a
single semiclassicality criterion, while reference sharpness
constitutes an additional consistency requirement.
The preceding comparisons separate three questions that are often
discussed together. First, validity concerns whether a given
semiclassical branch satisfies the approximations required for its
use as a controlled reference. Second, branch autonomy concerns
whether different semiclassical alternatives can be treated
independently when several of them coexist. Third, relational
interpretation concerns how evolution is described once a physical
reference has been identified.

The validity-domain construction developed here addresses primarily
the first question. Decoherence contributes to the second, while
relational and conditional-time formulations provide frameworks for
the third. Keeping these questions distinct prevents the existence,
autonomy, or relational interpretation of a semiclassical branch
from being used as substitutes for a quantitative test of its
validity as a temporal reference.

\revision{
A further limitation of the minisuperspace example
should be emphasized. In more realistic semiclassical
settings, a candidate gravitational reference may
represent a collective, coarse-grained variable built
from many microscopic quantum degrees of freedom.
Such a many-body or field-theoretic origin can control
its characteristic evolution scales, suppress relative
statistical fluctuations, and strongly suppress
interference between distinct coarse-grained histories.
These mechanisms cannot be exhibited within the
finite-dimensional homogeneous minisuperspace model
considered here. Accordingly, the FRW example is
intended only as a discriminating test of the logical
independence of the local WKB and Born--Oppenheimer
adiabatic conditions, rather than as a complete model
of the emergence and robustness of a semiclassical
clock. A field-theoretic extension of the
validity-domain construction would have to incorporate
these collective and decoherence effects explicitly.
}

\section{Discussion and Conclusions}
We have formulated the validity of semiclassical WKB time as a
problem of simultaneous control of distinct approximations on
gravitational configuration space. The central observation is that
the Hamilton--Jacobi flow can be defined before it is assigned a
temporal interpretation, allowing Born--Oppenheimer adiabaticity to
be tested directly through the configuration-space parameter
$\epsilon_{mn}^{\rm cfg}$.

For a single selected semiclassical branch, this leads to the
controlled temporal domain

\begin{equation}
\mathcal D_t
=
\mathcal D_{\rm WKB}
\cap
\mathcal D_{\rm ad}
\cap
\mathcal D_{\rm sharp},
\end{equation}

with an additional decoherence condition when several candidate
branches coexist coherently. The construction is therefore a
validity framework for the usual semiclassical temporal
parametrization rather than a new theory of time.
The closed FRW model provides a direct test of the independence of
the leading validity conditions. For a homogeneous massive scalar
mode, approaching the gravitational turning point from the
oscillatory region gives

\begin{equation}
\delta_{\rm WKB}
\longrightarrow
\infty,
\qquad
\epsilon_{n+2,n}^{\rm cfg}
\longrightarrow
0.
\end{equation}

Thus the subordinate sector becomes locally more adiabatic precisely
where the gravitational WKB approximation loses control. The
significance of this result is not the individual turning-point
behaviour of either parameter, but their opposite behaviour within
the same configuration-space region. It provides an explicit
demonstration, within the model, that

\begin{equation}
\mathcal D_{\rm ad}
\neq
\mathcal D_{\rm WKB}.
\end{equation}
This separation is robust under a uniform treatment of the turning
point. As shown in Appendix~\ref{app:airy}, the Airy solution
regularizes the divergent ordinary-WKB indicator to
$\delta_A(a_*)=2/\sqrt{3}=O(1)$, while the corresponding
configuration-space adiabatic parameter scales as
$\epsilon_{\rm cfg}^{A}(a_*)=O(M^{-1/3})$. The inequivalence of the
two validity conditions therefore does not rely on the singular
behavior of the ordinary WKB approximation.

Consequently, the existence of an oscillatory semiclassical branch
and the adiabaticity of the subordinate quantum dynamics are not
sufficient, separately, to establish a controlled semiclassical
temporal reference.
None of the individual ingredients entering this construction is
new. Gravitational WKB expansions, Born--Oppenheimer separation,
semiclassical Schr\"odinger dynamics, backreaction, decoherence, and
fluctuation criteria all have established precedents. The proposed
contribution lies instead in their organization around a specific
physical question: when can a semiclassical gravitational flow be
used as a quantitatively controlled temporal reference?

Within this organization, the configuration-space formulation of
adiabatic control plays a central role. It permits the suppression
of transitions between subordinate instantaneous channels to be
assessed before the Hamilton--Jacobi flow is parametrized by
$t_{\rm WKB}$. The usual time-dependent adiabatic criterion is then
recovered only after this configuration-space condition has been
defined. This removes the need to presuppose the semiclassical time
parameter in formulating the adiabatic test used to assess its domain
of applicability.
The resulting picture is stronger than a reformulation of the
standard semiclassical expansion. It shows that the emergence of a
Hamilton--Jacobi flow and the control of the quantum dynamics
defined relative to that flow are logically and quantitatively
distinct requirements. A semiclassical branch may therefore exist
without yet providing a reliable physical clock.

This suggests that semiclassical time should not be associated with
the mere appearance of an oscillatory WKB solution. Its controlled
use is instead a regional property of gravitational configuration
space, determined by the simultaneous validity of the approximations
that support the reference description. In this sense,
$\mathcal D_t$ provides a geometric criterion for where
semiclassical WKB time is physically reliable.
The present analysis nevertheless has a definite scope. The explicit
test has been carried out in a minisuperspace model, and the
sharpness condition has only been represented there by an
energy-fluctuation proxy. A field-theoretic extension would require
renormalized stress--energy correlations or an equivalent measure of
the induced fluctuations of the semiclassical geometry. Moreover,
the construction determines the validity of a candidate
semiclassical reference once a branch has been specified; it does
not provide a universal mechanism for selecting that branch.
The main conclusion is that semiclassical time is not controlled by
a single approximation criterion. A Hamilton--Jacobi flow may
exist, the subordinate quantum dynamics may even become increasingly
adiabatic, and yet the gravitational branch supporting that
description may simultaneously lose WKB validity. The FRW
turning-point analysis provides an explicit realization of this
separation.

This leads to a geometric characterization of semiclassical
temporal reliability:

\begin{equation}
\boxed{
\begin{gathered}
\text{controlled semiclassical time requires}\\
\text{overlap of the relevant validity domains}
\end{gathered}
}
\end{equation}

For a single selected branch, this overlap is
$\mathcal D_t=\mathcal D_{\rm WKB}\cap\mathcal D_{\rm ad}
\cap\mathcal D_{\rm sharp}$, with branch decoherence supplying an
additional condition when coherent alternatives coexist. The
emergence of an oscillatory WKB branch is therefore not the end of
the semiclassical-time problem: one must also determine where that
branch constitutes a quantitatively controlled physical reference.
\appendix

\section{Uniform Airy analysis of the gravitational turning point}
\label{app:airy}

The turning-point analysis of Sec.~V was formulated using the
ordinary local WKB expansion, for which
\begin{equation}
\delta_{\rm WKB}\rightarrow\infty,
\qquad
\epsilon_{\rm cfg}^{(n+2,n)}\rightarrow0
\qquad
(a\rightarrow a_*^+).
\end{equation}
Since the breakdown of the ordinary WKB approximation at a simple
turning point is standard, it is useful to verify whether the
inequivalence of the WKB and adiabatic control conditions survives
a uniform treatment of the gravitational wave function.

We therefore linearize the gravitational Wheeler--DeWitt equation
in the neighborhood of
\begin{equation}
a_*=H_\Lambda^{-1}.
\end{equation}
Neglecting subordinate terms and factor-ordering corrections, the
gravitational equation corresponding to Eq.~(27) may be written as
\begin{equation}
\frac{d^2\chi}{da^2}
+
\frac{36M^2}{\hbar^2}
a^2\left(H_\Lambda^2a^2-1\right)\chi
=
0.
\label{eq:airy_wdw}
\end{equation}

Introducing
\begin{equation}
x=a-a_*,
\end{equation}
one has
\begin{equation}
a^2\left(H_\Lambda^2a^2-1\right)
=
2a_*x+O(x^2).
\end{equation}
The leading turning-point equation is therefore
\begin{equation}
\frac{d^2\chi}{dx^2}
+
\lambda^3x\,\chi
=
0,
\label{eq:airy_local}
\end{equation}
where
\begin{equation}
\lambda^3
=
\frac{72M^2a_*}{\hbar^2}.
\label{eq:airy_lambda}
\end{equation}
Introducing
\begin{equation}
z=-\lambda x,
\end{equation}
Eq.~\eqref{eq:airy_local} becomes the standard Airy equation
\begin{equation}
\frac{d^2\chi}{dz^2}-z\chi=0.
\end{equation}

\subsection{Uniform solution and phase flow}

A real basis of solutions of Eq.~\eqref{eq:airy_local} is provided
by $Ai(z)$ and $Bi(z)$. In the oscillatory region $x>0$, corresponding
to $z<0$, a definite-flux combination can be chosen as
\begin{equation}
\chi_A(x)
=
Ai(-\lambda x)
+
i\,Bi(-\lambda x).
\label{eq:airy_branch}
\end{equation}

The sign is chosen to match the expanding branch used in Sec.~V,
for which the Hamilton--Jacobi momentum satisfies
\begin{equation}
p_a=\frac{dS_0}{da}<0.
\end{equation}

For $\lambda x\gg1$, the standard Airy asymptotics give
\begin{equation}
Ai(-\lambda x)+iBi(-\lambda x)
\propto
(\lambda x)^{-1/4}
\exp\left[
-i\frac{2}{3}(\lambda x)^{3/2}
\right],
\end{equation}
up to an irrelevant constant phase.

Writing the uniform solution as
\begin{equation}
\chi_A(x)
=
A_A(x)
\exp\left[
\frac{iM}{\hbar}S_A(x)
\right],
\end{equation}
the local phase momentum is
\begin{equation}
p_A(x)
\equiv
\frac{dS_A}{dx}
=
\frac{\hbar}{M}
\operatorname{Im}
\left(
\frac{\chi_A'}{\chi_A}
\right).
\end{equation}

Using the exact Airy Wronskian
\begin{equation}
Ai(z)Bi'(z)-Ai'(z)Bi(z)
=
\frac{1}{\pi},
\end{equation}
one obtains
\begin{equation}
p_A(x)
=
-
\frac{\hbar\lambda}{M\pi}
\frac{1}{
Ai^2(-\lambda x)+Bi^2(-\lambda x)
}.
\label{eq:airy_momentum}
\end{equation}

Unlike the leading Hamilton--Jacobi momentum, this expression remains
finite at the turning point.

As a check of the matching, for $\lambda x\gg1$,
\begin{equation}
Ai^2(-\lambda x)+Bi^2(-\lambda x)
\simeq
\frac{1}{\pi\sqrt{\lambda x}},
\end{equation}
so that
\begin{equation}
p_A(x)
\simeq
-
\frac{\hbar}{M}
\lambda^{3/2}\sqrt{x}.
\end{equation}
Using Eq.~\eqref{eq:airy_lambda},
\begin{equation}
\frac{\hbar}{M}\lambda^{3/2}
=
6\sqrt{2a_*},
\end{equation}
and therefore
\begin{equation}
p_A(x)
\simeq
-6\sqrt{2a_*x}.
\end{equation}

This is precisely the near-turning-point expansion of the
Hamilton--Jacobi momentum
\begin{equation}
p_a
=
-6a\sqrt{H_\Lambda^2a^2-1}
\end{equation}
used in Sec.~V. The uniform Airy branch therefore matches both the
sign and the normalization of the expanding WKB branch.
\subsection{Uniform WKB-control parameter}

We now apply the same local control indicator used in Sec.~V to the
phase gradient of the uniform Airy solution. We define
\begin{equation}
\delta_A(x)
\equiv
\frac{\hbar}{M}
\left|
\frac{p_A'(x)}{p_A^2(x)}
\right|.
\label{eq:airy_delta_def}
\end{equation}

Introducing
\begin{equation}
Q(z)
=
Ai^2(z)+Bi^2(z),
\end{equation}
Eq.~\eqref{eq:airy_momentum} gives
\begin{equation}
\delta_A(x)
=
2\pi
\left|
Ai(z)Ai'(z)+Bi(z)Bi'(z)
\right|,
\qquad
z=-\lambda x.
\label{eq:airy_delta_general}
\end{equation}

At the turning point $z=0$, the exact Airy values are
\begin{equation}
Ai(0)
=
\frac{1}{
3^{2/3}\Gamma(2/3)
},
\qquad
Bi(0)
=
\sqrt{3}\,Ai(0),
\end{equation}
and
\begin{equation}
Ai'(0)
=
-
\frac{1}{
3^{1/3}\Gamma(1/3)
},
\qquad
Bi'(0)
=
-\sqrt{3}\,Ai'(0).
\end{equation}

Using
\begin{equation}
\Gamma(1/3)\Gamma(2/3)
=
\frac{2\pi}{\sqrt{3}},
\end{equation}
one obtains
\begin{equation}
\delta_A(a_*)
=
\frac{2}{\sqrt{3}}
=
1.15470\ldots .
\label{eq:airy_delta_turning}
\end{equation}

Thus the divergence of the ordinary WKB indicator is removed by the
uniform Airy treatment, but the corresponding control parameter does
not become small:
\begin{equation}
\delta_A(a_*)=O(1).
\end{equation}

The turning point therefore remains outside a parametrically
controlled local WKB domain if the required criterion is
$\delta_{\rm WKB}\ll1$.  The uniform approximation regularizes the
wave function and its phase flow, but it does not turn the
gravitational turning point into a region of local WKB validity.
\subsection{Configuration-space adiabaticity with the uniform flow}

The uniform solution also allows us to examine the adiabatic control
parameter when the leading Hamilton--Jacobi flow is replaced by the
phase flow extracted from the Airy solution.

For the FRW model of Sec.~V,
\begin{equation}
V^a
=
G^{aa}p_a
=
-\frac{p_a}{6a}.
\end{equation}
Using Eq.~\eqref{eq:airy_momentum}, the uniform phase flow at the
turning point is therefore
\begin{equation}
V_A^a(a_*)
=
\frac{\hbar\lambda}{6Ma_*}
\frac{1}{
\pi[Ai^2(0)+Bi^2(0)]
}.
\end{equation}

Since
\begin{equation}
Ai^2(0)+Bi^2(0)
=
\frac{4}{
3^{4/3}\Gamma^2(2/3)
},
\end{equation}
one obtains
\begin{equation}
V_A^a(a_*)
=
\frac{\hbar\lambda}{6Ma_*}
\frac{
3^{4/3}\Gamma^2(2/3)
}{
4\pi
}.
\label{eq:airy_flow_turning}
\end{equation}

For the transition $n\rightarrow n+2$, Eq.~(13) together with
Eqs.~(38) and (39) gives
\begin{equation}
\epsilon_{\rm cfg}^{(n+2,n)}
=
\frac{3}{8am}
|V^a|
\sqrt{(n+1)(n+2)}.
\end{equation}
Hence, at the turning point,
\begin{equation}
\epsilon_{\rm cfg}^{A}(a_*)
=
\frac{
3^{4/3}\Gamma^2(2/3)
}{
64\pi
}
\frac{\hbar\lambda}{
Ma_*^2m
}
\sqrt{(n+1)(n+2)}.
\label{eq:airy_epsilon_exact}
\end{equation}

Using
\begin{equation}
\lambda
=
\left(
\frac{72M^2a_*}{\hbar^2}
\right)^{1/3},
\end{equation}
this may be written as
\begin{equation}
\epsilon_{\rm cfg}^{A}(a_*)
=
{\cal C}_A
\frac{
\hbar^{1/3}
}{
M^{1/3}m\,a_*^{5/3}
}
\sqrt{(n+1)(n+2)},
\label{eq:airy_epsilon_scaling}
\end{equation}
where
\begin{equation}
{\cal C}_A
=
\frac{
3^{4/3}\Gamma^2(2/3)
}{
64\pi
}
72^{1/3}
\end{equation}
is a purely numerical coefficient.

For fixed subordinate scales $m$, $a_*$ and fixed excitation level,
the semiclassical scaling is therefore
\begin{equation}
\epsilon_{\rm cfg}^{A}(a_*)
=
O(M^{-1/3}).
\label{eq:airy_epsilon_asymptotic}
\end{equation}

Thus the uniform treatment changes the exact limiting value
$\epsilon_{\rm cfg}\rightarrow0$ obtained from the leading
Hamilton--Jacobi flow into a finite but parametrically suppressed
quantity.  In the semiclassical regime,
\begin{equation}
\epsilon_{\rm cfg}^{A}(a_*)\ll1,
\end{equation}
while, from Eq.~\eqref{eq:airy_delta_turning},
\begin{equation}
\delta_A(a_*)=\frac{2}{\sqrt3}=O(1).
\end{equation}
The uniform turning-point treatment therefore replaces the ordinary
WKB behavior
\begin{equation}
\delta_{\rm WKB}\rightarrow\infty,
\qquad
\epsilon_{\rm cfg}^{(n+2,n)}\rightarrow0
\end{equation}
by
\begin{equation}
\delta_A(a_*)=O(1),
\qquad
\epsilon_{\rm cfg}^{A}(a_*)=O(M^{-1/3}).
\end{equation}

The formal WKB divergence is removed, but the two validity conditions
remain parametrically inequivalent.  In particular, Born--Oppenheimer
adiabaticity can remain well controlled while the gravitational WKB
criterion is not parametrically satisfied.  The conclusion
\begin{equation}
\mathcal D_{\rm ad}
\neq
\mathcal D_{\rm WKB}.
\end{equation}

therefore does not rely on the singular behavior of the ordinary WKB
approximation at the turning point.

The Airy analysis strengthens the role of the FRW turning point as a
discriminating test: the relevant result is not the divergence of one
control parameter and the exact vanishing of the other, but their
different parametric behavior under a uniform treatment of the same
configuration-space region.

\bibliography{references}

\end{document}